\documentstyle[prl,twocolumn,aps,epsf]{revtex}

\begin{document}
\twocolumn[
\hsize\textwidth\columnwidth\hsize\csname@twocolumnfalse\endcsname

\draft
\title{Frictional drag in dilute bilayer 2D hole systems}
\author{E. H. Hwang$^{(1)}$, S. \ Das Sarma$^{(1)}$, V. Braude$^{(2)}$, and Ady Stern$%
^{(2)}$}
\address{$^{(1)}$Condensed Matter Theory Center, Department of Physics, University of%
\\
Maryland, College Park, Maryland 20742-4111\\
$^{(2)}$ Department of Condensed Matter Physics, Weizmann Institute of\\
Science, Rehovot 76100, Israel }
\date{\today}
\maketitle

\begin{abstract}
We develop a theory for frictional drag between two 2D hole layers in a
dilute bilayer GaAs hole system, including effects of hole-hole and
hole-phonon interactions. Our calculations suggest significant enhancement
of hole drag transresistivity over the corresponding electron drag results.
This enhancement originates from the exchange induced renormalization of the
single layer compressibility and the strong dependence of single layer
conductivity on density. We also address the effect of hole-phonon
interaction on the drag temperature dependence. Our calculated results are
in reasonable quantitative agreement with recent experimental observations.

\end{abstract}
\vspace{0.5cm}
]

\newpage
Frictional drag measurements of transresistivity in modulation doped GaAs
electron bilayer systems have led to significant advances in our
understanding of density and temperature dependence of electron-electron and
electron-phonon interactions in 2D systems \cite{Rojo99}. Recent interest 
\cite{Noh98,Hill97,Flensberg,Swierkowski} has focussed on the role of
electron correlation effects on the drag resistivity, which should vary in a
systematic manner as a function of electron density and temperature. In
particular a recent experiment by Pillarisetty {\it et al}. \cite
{Pillarisetty} reports drag measurements in very low density and high
quality hole bilayers, where Coulomb interaction (i.e., correlation) effects
should be strong by virtue of the large GaAs hole effective mass ($%
m_{h}^{\ast }\approx 0.4m$ for holes compared with $m_{e}^{\ast }\approx
0.07m$ for electrons where $m$ is the vacuum electron mass) and very low
hole density (hole density $p=10^{10}-10^{11}cm^{-2}$ in Ref. %
\onlinecite{Pillarisetty}, whereas typical electron densities used bilayer
drag measurements have been comparatively high, $>10^{11}cm^{-2}$). In terms
of the dimensionless interaction strength parameter $r_{s}\equiv (p\pi
)^{-1/2}m^{\ast }e^{2}/(\hbar ^{2}\kappa )$ where $\kappa $ is the
background dielectric constant, which measures the ratio of the potential
energy to the kinetic energy in the interacting hole system, Ref. %
\onlinecite{Pillarisetty} explores the strongly correlated regime of $%
r_{s}\approx 20-40$ whereas the earlier electron drag experiments explored
the weak coupling regime of $r_{s}<3$.

The experimental findings of Ref. \cite{Pillarisetty} were striking, and
motivated our work. They are: $a.$ the low density hole drag is $2-3$ orders
of magnitude larger than the corresponding electron drag results published
in the literature \cite{Rojo99,Noh98,Hill97,Gramila93}; $b.$ There are some
small (but systematic) deviations of the observed low density hole drag
resistivity from the expected $\rho \sim T^{2}$ Fermi liquid behavior. The
low density data of Ref. \onlinecite{Pillarisetty} seem to better fit a $%
T^{2.5}$ behavior at low temperatures; $c.$ The observed $\rho /T^{2}$
behavior in Ref. \onlinecite{Pillarisetty}, plotted as a function of $T$ for
various bilayer hole densities, is qualitatively similar to the
corresponding electron drag results in the sense that $\rho /T^{2}$ at a
fixed density shows a peak at some temperature $T_{p}$ which decreases with
decreasing density -- the peak in $\rho /T^{2}$ as a function of $T$ at the
lowest hole density $p=10^{10}cm^{-2}$ is very sharp; $d.$ For bilayers with
unequal hole densities, the drag resistivity at a fixed temperature plotted
as a function of the density ratio $p_{1}/p_{2}$ decreases monotonically and
does not exhibit a peak at the balance point as it does in the corresponding
electron case. This peak is believed to arise from the $2k_{F}$ phonon
scattering.

These experimental findings become particularly interesting due to their
possible relation to the collection of transport anomalies in 2D systems
referred to as the 2D metal-insulator transition (2D MIT) phenomena (The
samples of Ref. \onlinecite{Pillarisetty} exhibit 2D MIT in each layer at $%
p\approx 8.5\times 10^{9}cm^{-2}$). These anomalies, observed largely in the
large $r_{s}$ regime, have raised doubts regarding the applicability of
Fermi liquid theory to two dimensional systems of charges (electrons or
holes) at the large $r_{s\text{ }}$regime, and this question has been widely
debated in the literature (see \cite{Abrahams} for a review). In that
context, then, it is particularly important to examine whether drag
measurements in the large $r_{s}$ regime may be understood within the Fermi
liquid framework.

In this paper we attempt, and largely succeed, to interpret the hole drag
data of Ref. \onlinecite{Pillarisetty} within a Fermi liquid approach. The
important inputs to our theory are a perturbative expression to the drag
resistivity, based on a perturbative treatement of inter-layer hole-hole
scattering rate, a Hubbard approximation for the single layer polarization
operator, and the experimentally measured density dependence of the single
layer conductivity. Inputs of more minor significance are form factors that
account for the finite thickness of the two layers and hole-phonon
interaction parameters.

Our starting point for the calculation of the drag resistivity is the
following expression: 
\begin{equation}
\rho =\frac{\beta }{\sigma _{1}\sigma _{2}}\frac{d\sigma _{1}}{dp_{1}}\frac{%
d\sigma _{2}}{dp_{2}}\int \frac{q^{2}d^{2}q}{(2\pi )^{2}}\frac{d\omega }{%
2\pi }\frac{F_{1}(q,\omega )F_{2}(q,\omega )}{\sinh ^{2}(\beta \omega /2)},
\label{eq:rhod}
\end{equation}
where $F_{i}(q,\omega )={%
\mathop{\rm Im}%
}\Pi _{ii}(q,\omega )|u_{12}^{sc}(q,\omega )|$. In Eq. (1), $\beta =1/T$ is
the inverse temperature (we use units such that $k_{B}=\hbar =2e=1$, except
in final formulas); $p_{1,2}$ are the hole densities in layers 1 and 2; $%
\sigma _{1,2}$ are the conductivities of each layer; $q$ is the 2D wave
vector in the layer; $\Pi _{11}$/$\Pi _{22}$ are the irreducible hole
polarizabilities in each layer; and $u_{12}^{sc}$ is the dynamically
screened effective interlayer interaction. Several comments are in place
regarding this expression: first, when the drag resistivity is derived from
the Boltzmann equation (\cite{Jauho,Bonsager}, and see also \cite{Dassarma}%
), the approximation $\frac{d\sigma _{i}}{dp_{i}}\approx \frac{\sigma _{i}}{%
p_{i}}$ is being made. This approximation is valid for well conducting
layers, but becomes invalid at the low densities relevant here (see below).
The need to replace $\frac{\sigma _{i}}{p_{i}}$ by $\frac{d\sigma _{i}}{%
dp_{i}}$ is discussed in Ref. \cite{stern}. Second, the dynamically screened
interlayer interaction $u_{12}^{sc}$ satisfies a matrix Dyson equation, 
\begin{equation}
{\bf u}(q,\omega )=\left[ {\bf 1}-{\bf v}_{t}(q,\omega )\Pi (q,\omega )%
\right] ^{-1}{\bf v}_{t}(q,\omega ).
\end{equation}
We include in the bare interaction ${\bf v}_{t}(q,\omega )={\bf v}^{c}(q)+%
{\bf v}^{ph}(q,\omega )$ both the hole-hole direct Coulomb interaction and
the phonon mediated interaction ${\bf v}^{ph}$, which includes the acoustic
phonon propagator and the appropriate hole-phonon interaction matrix element 
\cite{Hwang01p}. The phonon mediated interaction includes both deformation
potential and piezoelectric couplings between the holes and the acoustic
phonons using the standard hole-phonon interaction parameters for GaAs \cite
{Bonsager}. Following the common approximation we assume the inter-layer
polarization operators $\Pi _{12},\Pi _{21\text{ }}$to be zero.

When Eq. (\ref{eq:rhod}) is used to analyze experiments carried out on
identical layers in the commonly explored regime,\ of high density and low
temperature, it yields, 
\begin{equation}
\rho _{D}=\frac{\zeta (3)\pi }{32}\frac{h}{e^{2}}\left( \frac{k_{B}T\kappa }{%
e^{2}k_{F}^{3}d^{2}}\right) ^{2}  \label{boltzmann-estimate}
\end{equation}
with $\zeta$ being the Riemann zeta function. The assumptions involved in
getting from Eq. (\ref{eq:rhod}) to Eq. (\ref{boltzmann-estimate}) are large
inter-layer separation ($k_{F}d\gg 1,$ $q_{TF}d\gg 1$, with $k_{F}$ being
the Fermi wave vector and $q_{TF}$ being the Thomas Fermi screening wave
vector), Drude relation between conductivity and density ($\sigma
_{i}\propto p_{i}$), and a Random Phase Approximation (RPA) in which $\Pi
_{ii}$ is replaced by its value for non-interacting electrons. The
contribution of electron-phonon interaction is negligible in this regime.
Eq. (\ref{boltzmann-estimate}) is typically smaller than the experimentally
measured value, by a factor of $2-5.$ For the experiment of Ref. 
\onlinecite{Pillarisetty}, 
the disagreement is much larger, getting as large as a
factor of $500.$

Our calculation differs from that leading to Eq. (\ref{boltzmann-estimate})
in five points, all of them leading to an increase of the drag resistivity.

First, we use the Hubbard approximation to calculate the polarization
operators $\Pi _{ii}$. Within this approximation, 
\begin{equation}
\Pi _{ii}(q,\omega )=[1+v_{i}^{c}(q)\Pi _{ii}^{0}(q,\omega )G(q)]^{-1}\Pi
_{ii}^{0}(q,\omega ),  \label{hubbard-approx}
\end{equation}
where $\Pi _{ii}^{0}(q,\omega )$ is the polarization operator for
non-interacting electrons, and $G(q)=q/2\sqrt{q^{2}+k_{F}^{2}}$ is the local
field correction. For small $q$ the Hubbard approximation amounts to the
introduction of a Landau parameter $f_{0}=-v_{i}^{c}(q)G(q)\approx \frac{\pi
e^{2}}{k_{F}}$, making the inverse compressibility of each layer $\frac{%
\partial \mu }{\partial n}=\frac{2\pi \hbar ^{2}}{m}-\frac{\pi e^{2}}{k_{F}}%
. $ This approximation is then consistent with measurements of the
compressibility that yield negative values.
In general, the
effect of the Hubbard approximation is to decrease the bare intralayer
interaction by a factor of $1-G(q)$. The fact that the exchange-driven local
field correction, $G(q)$, affects only the intra-layer interaction is what
makes its effect significant: the interaction potential between
anti-symmetric charge densities, $v_{11}^{c}(q)(1-G(q))-v_{12}^{c}(q)$, is
modified from $2\pi e^{2}d/\kappa $ to $\frac{2\pi e^{2}}{\kappa }(d-\frac{1%
}{2k_{F}})$. This increases the drag resistivity, and in the limit of $%
k_{F}d\gg 1$, $q_{TF}d \gg 1$, 
Eq. (\ref{boltzmann-estimate}) is replaced by 
\begin{equation}
\rho _{D}=\frac{\zeta (3)\pi }{32}\frac{h}{e^{2}}\left( \frac{k_{B}T\kappa }{%
e^{2}k_{F}^{3}d^{2}}\right) ^{2}\left( \frac{2k_{F}d}{2k_{F}d-1}\right) ^{4}.
\label{hubbard-estimate}
\end{equation}
We note by passing that even for densities that are not very low, the fourth
power of the last term in Eq. (\ref{hubbard-estimate}) makes the correction
quite significant. In the low density regime we consider, the approximation 
Eq. (\ref{hubbard-estimate}) is not valid, since scattreing events with large
momentum transfer have a significant contribution. Our calculations below
show that in that regime the local field correction increases the drag
resistivity by about an order of magnitude.

Second, we do not use the approximation $\sigma _{i}\propto p_{i}.$ Rather,
we use measured values of $\sigma (p),$ provided to us by Pillarisetty {\it %
et al}. \cite{Tsui}, 
to extract $\frac{d\sigma _{i}}{dp_{i}}$ for use in our calculation
of the drag resistivity. At the lowest density measured $\left( \frac{%
d\sigma _{i}}{dp_{i}}\frac{p_{i}}{\sigma _{i}}\right) ^{2}\simeq 10-20$, so
this correction increases the drag resistivity by about an order of
magnitude. These measurements of $\sigma (p)$ are limited, at this stage, to
the range of densities between $0.72-2.15\cdot 10^{10}cm^{-2}$, and to the
temperature range $T<0.5 K$. While this correction deacreases with
increasing density and temperature, it is still between $3-5$ at the edge of
the measured range. Our calculation for the range where experimental values
of $\left( \frac{d\sigma _{i}}{dp_{i}}\frac{p_{i}}{\sigma _{i}}\right) ^{2}$
are available is presented in Fig. 1. The other figures do not include
this factor, for lack of availability of experimental data.

Third, the assumption $k_{F}d>>1$, implying that the relevant momentum
exchange is much smaller than $k_{F}$, is not valid at the low density
regime. Consequently, there is an important contribution due to large
momentum transfer, $\hbar q\approx 2\hbar k_{F}$ scattering. Taking this
contribution into account (by a numerical integration of Eq. (\ref{eq:rhod}%
)) leads to an increase of the drag by another factor of $\sim 2$ as
compared to the Boltzmann result.

Fourth, the introduction of finite thickness form factors \cite{flensberg}
(see to the Coulomb and phonon interaction)

\begin{figure}[tbp]
\epsfysize=3.3in
\centerline{\epsffile{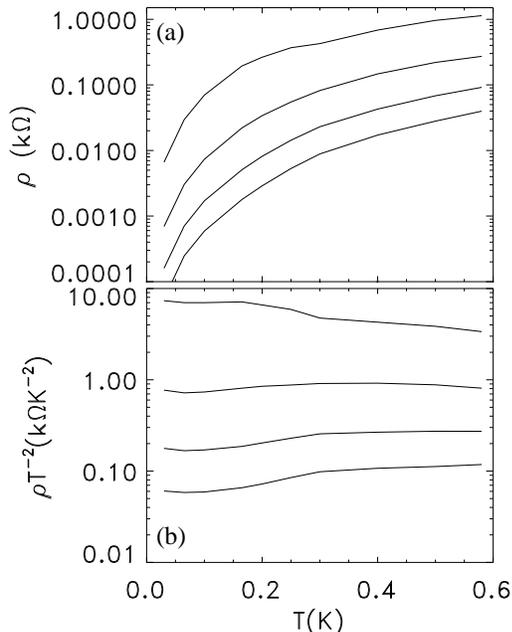}}
\caption{The drag resistivity (a) and $\protect\rho (T)/T^{2}$ (b)
as a function of temperature for various hole
densities ($p=1.0,$ 1.5, 2.0, 2.5 $\times 10^{10}cm^{-2}$, from top
to bottom) calculated with all five correction factors as explained 
in the text. 
Throughout this paper we use a hole bilayer system with the layer
separation of $d=300\AA $ and the well width of $a=150\AA .$}
\end{figure}
\noindent
effectively decreases $d$, and increases the
drag resistivity by about a factor $2.$

And fifth, we include the phonon contribution to the interaction in addition
to the Coulomb one. This changes the result by less then $50\%$.

Combining all these five factors, we are able to account for most of the
results of the measurements. We account for the very large increase of drag,
as compared to measurements of electronic systems. Our Fig. 1 is in good
quantitative agreement with the measured data, to within a factor of $2$.
This type of agreement is similar to what is obtained in the small $r_{s}$
limit. Our analysis yields a leading quadratic temperature dependence of the
drag in the limit $T\longrightarrow 0$ (at least as long as the conductivity 
$\sigma $ is temperature independent in that limit). However, our numerical
integration of Eq. \ref{eq:rhod}, as presented in Fig. 2, indicates that
even at the lowest measured temperatures the drag resistivity does not
follow a $T^{2}$ dependence. There are a number of reasons for that. First,
at the low densities used in Ref. 
\onlinecite{Pillarisetty} [shown as 
the inset in our Fig. 2 to be compared with inset (b) in 
Fig. 3 of Ref. \onlinecite{Pillarisetty}] the Fermi energy ($1-2K)$ is
not much larger than the measurement temperature range. Second, even well
below $1K$ \ the phonon contribution to drag in the low density hole
bilayers is quite substantial (in contrast to electron systems where the
phonon contribution is typically a factor of $10^{3}$ smaller for small
layer separations 
of $d=300\AA $ or so used in Ref. \onlinecite{Pillarisetty}%
). As such we believe that the experimental departure from the $T^{2}$
behavior 

\begin{figure}[tbp]
\epsfysize=2.1in
\centerline{\epsffile{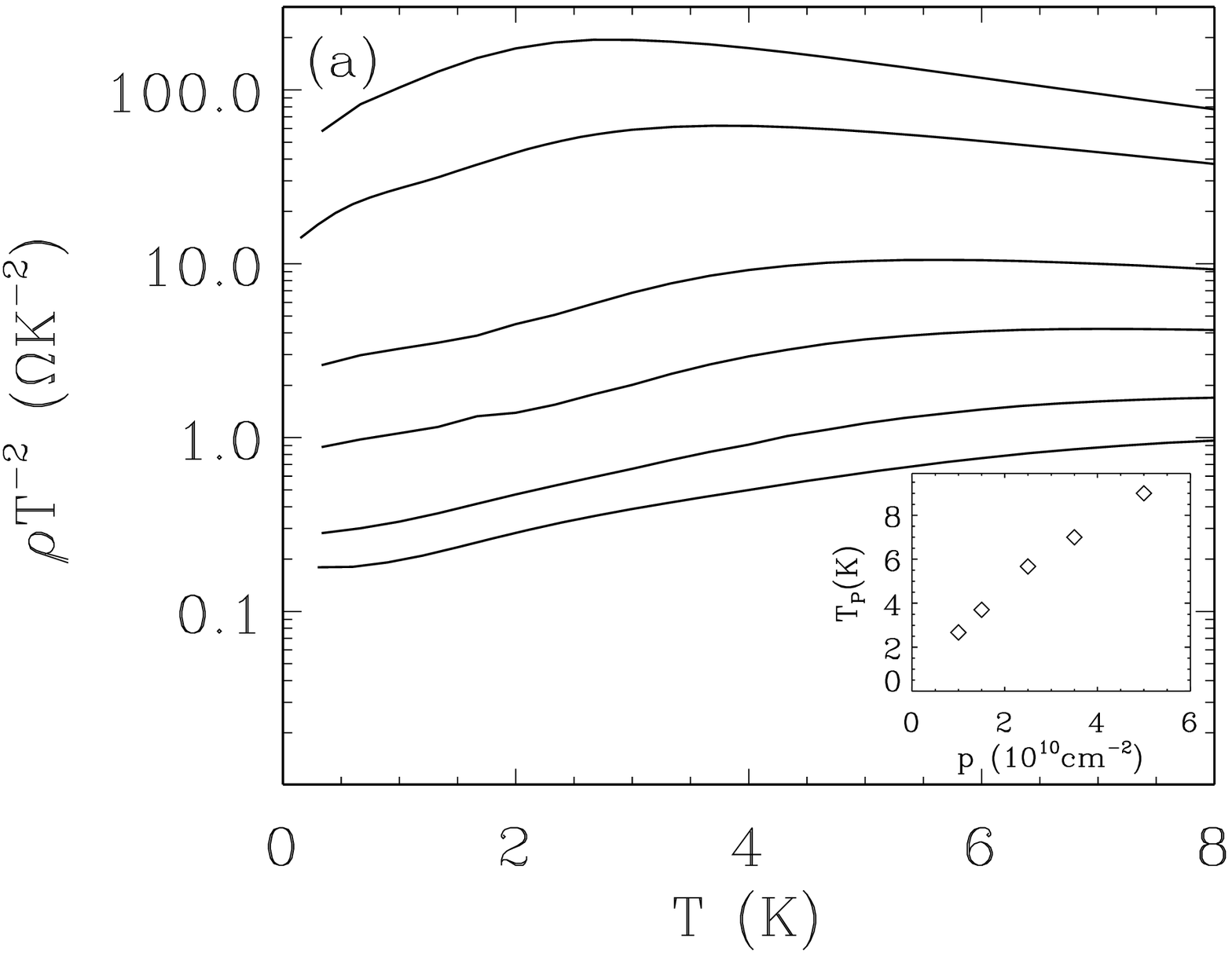}}
\epsfysize=2.1in
\centerline{\epsffile{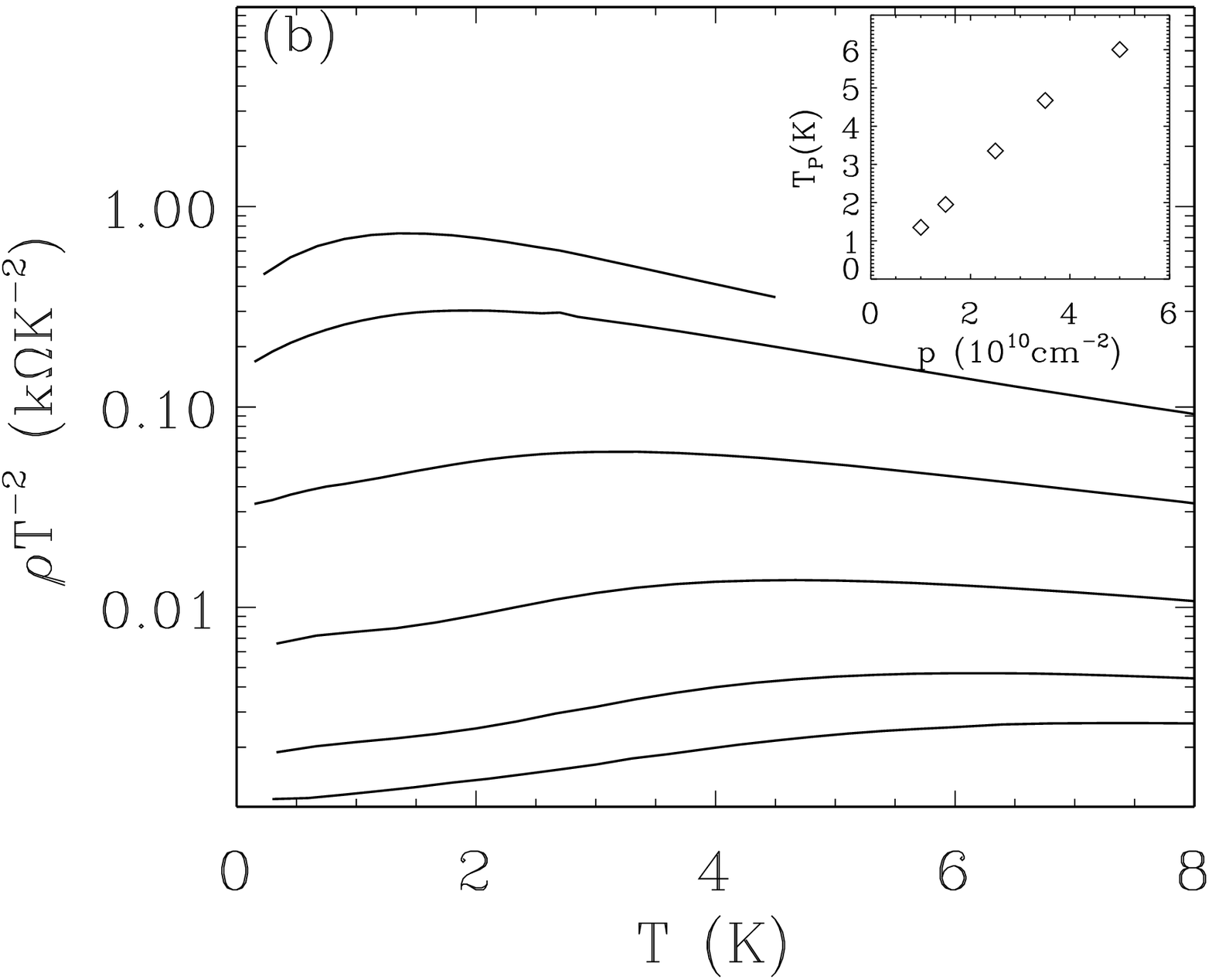}}
\caption{The calculated total hole frictional drag $\protect\rho (T)/T^{2}$
as a function of $T$ for various hole densities ($p=1.0,$ 1.5, 2.5, 3.5,
5.0, $7.0\times 10^{10}cm^{-2}$, from top to bottom) within (a) RPA
dynamical screening theory, and (b) HA. In insets we show the variation in
the calculated peak temperature $T_{p}$ as a function of the hole density 
$p$.}
\end{figure}
\noindent
reported in Ref. \onlinecite{Pillarisetty} is essentially a
manifestation of the fact that phonon effects remain significant in the
experiments, and the asymptotic $T^{2}$ regime is hard to reach in hole
systems. We find that our calculated $\rho (T)$ at low temperature is well
approximated by a $T^{2.4}$ behavior for $p=2.0\times 10^{10}cm^{-2}$ and
the exponent increases as the hole density decreases. In Fig. 4 we show our
calculated contributions to the hole drag resistivity from individual
hole-hole and hole-phonon interactions as compared with the corresponding
electron case. The importance of phonon effects to the hole drag
transresistivity is manifestly evident in Fig. 4.

In Fig. 3 we qualitatively ``explain'' the particularly anomalous feature of
the data in Ref. 
\onlinecite{Pillarisetty} [shown as 
the inset in our Fig. 2 to be compared with inset (b) in 
Fig. 3 of Ref. \onlinecite{Pillarisetty}], i.e., the non-existence of a peak
in $\rho /T^{2}$ as a function of the density ratio $p_{1}/p_{2}.$ In
qualitative agreement with Ref. \onlinecite{Pillarisetty} the calculated
drag resistivity at a fixed temperature decreases monotonically as a
function of the hole density ratio $p_{1}/p_{2}$ without showing any
phonon-induced peak at the balance point $p_{1}=p_{2},$ 
as has been observed
in corresponding electron bilayer experiments \cite{Gramila93}. This peak
arises from the sharp 2D Fermi surfaces in the two electron layers which,
when perfectly 

\begin{figure}[tbp]
\epsfysize=2.1in
\centerline{\epsffile{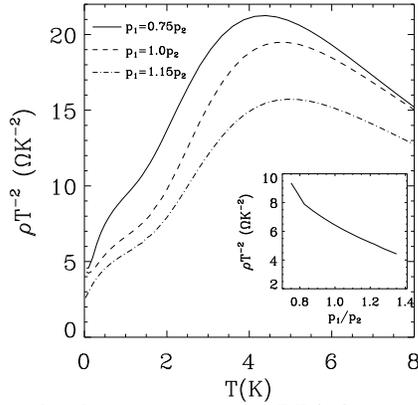}}
\caption{ The drag resistivity in RPA for various density ratio ($p_1/p_2$,
where $p_1 = p_{drive}$ and $p_2 = p_{drag}$) with the fixed drag layer
density ($p_2=2.0\times 10^{10} cm^{-2}$). Inset shows the drag resistivity
as a function of a density ratio at $T=1.0K$. }
\end{figure}
\noindent
matched at the balance point $p_{1}=p_{2}$, leads to enhanced
phonon scattering, leading to the peak resistivity at $p_{1}=p_{2}$. No such
peak exists in the hole bilayer case because of the small Fermi temperature
in the hole case, $T_{F}=1.4K$, and the large Bloch-Gr\"{u}neisen
temperature, $T_{BG}=2.8K$, (for $p=2\times 10^{10}cm^{-2}$). Therefore
typical $T/T_{F}$ is rather large in the hole case, leading to thermally
broadened Fermi surfaces in the low density hole bilayers of Ref. %
\onlinecite{Pillarisetty} which cannot exhibit any sharp Fermi surface
effects, and consequently the so-called ``$2k_{F}$ phonon peak'' 
\cite{Gramila93} arising from the 
Fermi surface matching is absent. The peak in $%
\rho _{D}/T^{2},$ appearing in Fig. 3, takes place at high temperatures,
above the Fermi energy, and is unlikely to be related to the experimental
peak that 
appears around $0.5 K.$ We believe it may be related to the two
opposing temperature dependences of the hole-hole 

\begin{figure}[tbp]
\epsfysize=3.in
\centerline{\epsffile{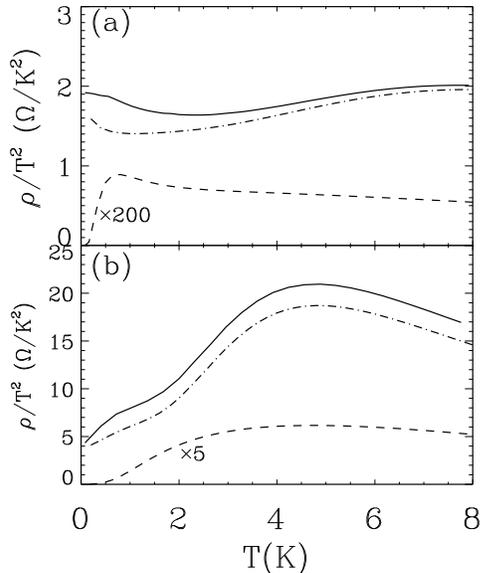}}
\caption{ (a) The RPA-calculated electron drag resistivities from
electron-electron $v^{c}$ (dot-dashed), electron-phonon $v^{ph}$ (dashed),
and total interaction $v_{t}$ (solid) for $n=2.0\times 10^{10}cm^{-2}$. (b)
The same for the hole drag resistivity. }
\end{figure}

\noindent
scattering rate on one
hand and the $\left( \frac{d\sigma _{i}}{dp_{i}}\frac{p_{i}}{\sigma _{i}}%
\right) ^{2}$ factor on the other hand.

In conclusion we have developed a theory for frictional interlayer drag
resistivity in low density hole bilayers including both hole-hole and
hole-phonon interaction effects. We explain the observed \cite{Pillarisetty}
dramatic deviation of drag resistivity from an RPA-based Boltzmann equation
calculation as arising from several factors, out of which the most important
ones are the dependence of the single-layer resistivity on density and the
exchange induced renormalization of the single-layer polarization operator.
Our calculation explains the large increase in the magnitude of drag and the
absence of phonon-induced $2k_{F}$ peak in the drag between matched
densities. We also obtain reasonable qualitative understaning of the
temperature dependence of the drag resistivity. Our calculation are all done
within a perturbative Fermi-liquid based approach.

The authors are grateful to D. C. Tsui for discussing the experimental
results of Ref. \onlinecite{Pillarisetty} prior to publication, and to R.
Pillarisetty and H. Noh for discussing unpublished measurements of
single layer resistivity. This work is supported by US-ONR, ARDA, DARPA,
US-Israel BSF and the Israel Science Foundation.


\vspace{-.5cm}

\begin{references}

\vspace{-1.5cm}

\bibitem{Rojo99}  A. G. Rojo, J. Phys.: Condens. Matt. {\bf 11}, R31 (1999);
references therein.

\bibitem{Noh98}  H. Noh {\it et al.}, \prb {\bf 58}, 12621 (1998).

\bibitem{Hill97}  N. P. R. Hill {\it et al.}, \prl {\bf 78}, 2204 (1997).

\bibitem{Flensberg}  K. Flensberg {\it et al.}, \prb {\bf 52}, 14761 (1995).

\bibitem{Swierkowski}  L. Swierkowski {\it et al}., \prb {\bf 55}, 2280
(1997); U. Sivan, P. M. Solomon, and H. Shtrikman, Phys. Rev. Lett. 68,
1196-1199 (1992)

\bibitem{Pillarisetty}  R. Pillarisetty {\it et al}., cond-mat/0202077
(submitted to \prl).

\bibitem{Gramila93}  T. J. Gramila {\it et al.}, \prb {\bf 47}, 12957
(1993); H. Rubel {\it et al}., Sem. Sci. Tech. {\bf 10}, 1229 (1995); H. Noh 
{\it et al}., \prb {\bf 59}, 13114 (1999).


\bibitem{Abrahams}  For recent reviews see, for example, E. Abrahams {\it et
al.}, Rev. Mod. Phys. {\bf 73}, 251 (2001); B. L. Altshuler {\it et al.},
Physica E {\bf 9}, 209 (2001).


\bibitem{Jauho}  A. P. Jauho and H. Smith, \prb {\bf 47}, 4420 (1993); L.
Zheng and A. H. MacDonald, \prb {\bf 48}, 8203 (1993).

\bibitem{Bonsager}  M. Bonsager {\it et al}., \prb {\bf 57}, 7085 (1998).

\bibitem{Dassarma}  S. Das Sarma {\it et al.}, \prb {\bf 41}, 3561 (1990);
J. Senna and S. Das Sarma, Solid State Commun. {\bf 64}, 1397 (1987).


\bibitem{stern}  F. von Oppen, S. H. Simon, and A. Stern, Phys. Rev. Lett.
87, 106803 (2001); B. N. Narozhny, I. L. Aleiner, and A. Stern, Phys. Rev.
Lett. 86, 3610-3613 (2001).

\bibitem{Hwang01p}  E. H. Hwang and S. Das Sarma, \prb {\bf 63}, 233201
(2001).

\bibitem{Tsui} R. Pillarisetty, H. Noh, and D. C. Tsui, private communication.

\bibitem{flensberg}  K. Flensberg and B.Y.K. Hu, Phys. Rev. B 52, 14796
(1995) (see Eq. 22).






\end{references}
\end{document}